\documentclass[sigconf]{acmart}

\usepackage{graphicx}
\usepackage{textcomp}
\usepackage{pifont}
\usepackage{xcolor}
\usepackage{soul}
\usepackage{booktabs}
\usepackage{multirow}
\usepackage{float}
\usepackage{url}
\usepackage{balance}
\usepackage{microtype}
\usepackage[numbered]{bookmark}
\usepackage{seqsplit}
\usepackage{newunicodechar,graphicx}
\usepackage[utf8]{inputenc}
\usepackage{pgf-pie}  
\usepackage{balance} 
\usepackage{algorithmic}
\usepackage{array}
\usepackage{tabularx} 
\usepackage{ragged2e}
\usepackage{standalone}
\usepackage{enumitem}
\setlist[itemize,enumerate]{noitemsep, topsep=0pt, leftmargin=1.0em}
\setlist{nolistsep}
\usepackage[utf8]{inputenc}
\usepackage{newunicodechar,graphicx}
\DeclareRobustCommand{\okina}{%
  \raisebox{\dimexpr\fontcharht\font`A-\height}{%
    \scalebox{0.8}{`}%
  }%
}
\newunicodechar{ʻ}{\okina}

\copyrightyear{2026}
\acmYear{2026}
\setcopyright{cc}
\setcctype{by}
\acmConference[CHASE'26]{19th International Conference International Workshop on Cooperative and Human Aspects of Software }{April 12--18, 2026}{Rio de Janeiro, Brazil}
\acmBooktitle{19th International Conference International Workshop on Cooperative and Human Aspects of Software (CHASE'26), April 12--18, 2026, Rio de Janeiro, Brazil}
\acmPrice{}
\acmDOI{10.1145/3794860.3794908}
\acmISBN{979-8-4007-2484-8/2026/04}

\begin{document}

\title{Understanding npm Developers’ Practices, Challenges, and Recommendations for Secure Package Development}

\author{Anthony Peruma}
\email{peruma@hawaii.edu}
\orcid{0000-0003-2585-657X}
\affiliation{%
  \institution{University of Hawai‘i at Mānoa}
  \state{Hawai‘i}
  \country{USA}
}

\author{Truman Choy}
\email{choytr@hawaii.edu}
\orcid{0009-0002-6336-7873}
\affiliation{%
  \institution{University of Hawai‘i at Mānoa}
  \state{Hawai‘i}
  \country{USA}
}

\author{Gerald Lee}
\email{glee25@hawaii.edu}
\orcid{0009-0001-4569-3950}
\affiliation{%
  \institution{University of Hawai‘i at Mānoa}
  \state{Hawai‘i}
  \country{USA}
}

\author{Italo De Oliveira Santos}
\email{isantos3@hawaii.edu}
\orcid{0000-0002-7545-6104}
\affiliation{%
  \institution{University of Hawai‘i at Mānoa}
  \state{Hawai‘i}
  \country{USA}
}

\newcommand{\SurveyParticipantCount}{75 }

\newcommand{\RQA}{\textbf{RQ1}: How do npm developers perceive security for their packages?}

\newcommand{\RQB}{\textbf{RQ2}: What security practices and tools do npm developers leverage in building and maintaining their packages?}

\newcommand{\RQC}{\textbf{RQ3}: What barriers hinder the secure development and maintenance of npm packages?}

\newcommand{\RQD}{\textbf{RQ4}: What improvements should be prioritized to strengthen security for npm packages?}

\begin{abstract}
  \textbf{\textit{Background:}} The Node Package Manager (npm) ecosystem plays a vital role in modern software development by providing a vast repository of packages and tools that developers can use to implement their software systems. However, recent vulnerabilities in third-party packages have led to serious security breaches, compromising the integrity of applications that depend on them. \textbf{\textit{Objective:}} This study investigates how npm package developers perceive and handle security in their work. We examined developers' understanding of security risks, the practices and tools they use, the barriers to stronger security measures, and their suggestions for improving the npm ecosystem's security. \textbf{\textit{Method:}} We conducted an online survey with 75 npm package developers and undertook a mixed-methods approach to analyzing their responses. \textbf{\textit{Results:}} While developers prioritize security, they perceive their packages as only moderately secure, with concerns about supply chain attacks, dependency vulnerabilities, and malicious code. Only 40\% are satisfied with the current npm security tools due to issues such as alert fatigue. Automated methods such as two-factor authentication and npm audit are favored over code reviews. Many drop dependencies due to abandonment or vulnerabilities, and typically respond to vulnerabilities in their packages by quickly releasing patches. Key barriers include time constraints and high false-positive rates. To improve npm security, developers seek better detection tools, clearer documentation, stronger account protections, and more education initiatives.
  \textbf{\textit{Conclusion:}} Our findings will benefit npm package contributors and maintainers by highlighting prevalent security challenges and promoting discussions on best practices 
  to strengthen security and trustworthiness within the npm landscape.
\end{abstract}




\maketitle

\section{Introduction}
\label{sec:introduction}

Modern software development increasingly relies on third-party packages that provide pre-built functionality and libraries to accelerate application development~\cite{Chowdhury2022}. Within this landscape, the Node Package Manager (npm)\footnote{\url{https://www.npmjs.com/}}, the
 default package manager for Node.js\footnote{\url{https://nodejs.org/}}, has become a key element of modern software ecosystems, facilitating developers in managing, distributing, and reusing code in an efficient and scalable way. With its vast registry of over 2 million packages and more than a billion monthly downloads~\cite{npm-download}, npm enables developers to quickly use pre-built solutions, significantly reducing the time and effort required to build their applications~\cite{abdalkareem2017developers}. However, this scale and interdependency introduce significant security challenges~\cite{zimmer}. Vulnerabilities in third-party packages can cascade through the software supply chain, leading to severe security breaches and undermining trust in the ecosystem~\cite{Ohm2020}.
 
Recent high-profile security incidents in the npm ecosystem highlight the severity of this threat. In September 2025, a major npm attack compromised 18 popular libraries, injecting malicious code that hijacked cryptocurrency wallets and manipulated blockchain transactions. The attack began with a phishing scheme targeting the maintainer’s npm account. Once access was gained, the attacker published malicious versions of widely downloaded packages~\cite{npmSecurityExample}. Such events underscore the socio-technical nature of software supply chain security, where technical measures and human behavior must be aligned to create a robust defense~\cite{williams2025research}. 

Although prior research has explored npm from various perspectives, including package abandonment~\cite{miller2024understanding}, maintainability and security analysis~\cite{chatzidimitriou2018npm}, non-coding contributions~\cite{canovas2022analysis}, supply chain dependencies and popularity~\cite{dey2018software,zerouali2019diversity}, pull request acceptance factors~\cite{dey2020effect}, ecosystem growth and dependency evolution~\cite{wittern2016look}, the prevalence of trivial packages~\cite{abdalkareem2017developers}, and vulnerabilities across ecosystems~\cite{zerouali2022impact,alfadel2023empirical}, there is limited understanding of the specific security practices and challenges faced by npm package developers.

Developers are central to software development and play a crucial role in shaping the security of the systems they build. Therefore, to achieve effective software security, it is important to understand their practices, priorities, and challenges, often through qualitative empirical studies \cite{Gutfleisch2022,Lopez2023}. To this end, we adopt a socio-technical approach, examining both the technical aspects and human factors of npm package security, which provides richer insights than purely technical analyses \cite{Storey2020, Stol2018}. By directly engaging with maintainers and contributors of real-world npm packages, our study aims to gather nuanced perspectives that accurately reflect the practices and challenges involved in achieving effective package security.

\subsection{Goal and Research Questions}
\label{sec:goal}
In this study, we conduct a comprehensive survey of npm package maintainers and contributors, using a combination of quantitative and qualitative questions. The goal of this research is to develop an evidence-based understanding of how npm package developers\footnote{We use the term npm package developers as an umbrella term encompassing maintainers and contributors who publish, maintain, or contribute to npm packages.} approach security by gathering \textit{insights directly from practitioners on how they perceive package security, what practices and tools they use, the challenges they face, and the improvements they recommend}.  
Our findings aim to advance secure software development in open source software (OSS) package management, offering actionable insights for tooling, education, and ecosystem support. We aim to answer the following research questions (RQs):

\vspace{1mm}
\noindent\textbf{\RQA} - Developers' security practices are shaped by how they weigh security against other priorities~\cite{Assal2025, Peruma2024}. This RQ investigates how npm package developers prioritize security, aiming to reveal gaps between perceptions and actual threats to effectively guide security prioritization and tool design.

\vspace{1mm}
\noindent\textbf{\RQB} - 
As security practices evolve, developers must adopt tools and best practices to mitigate risks~\cite{Assal2025}. 
This RQ explores how npm package developers utilize security tools, identifying common and rare practices, gaps hindering adoption, deviations from guidelines, and practices needing redesign or better advocacy.

\vspace{1mm}
\noindent\textbf{\RQC} - 
Understanding software security challenges is critical for improving practices and mitigating vulnerabilities~\cite{Ryan2023}. This RQ identifies technical, organizational, and human barriers faced by npm package developers, providing evidence to guide interventions and tool improvements.

\vspace{1mm}
\noindent\textbf{\RQD} - 
Software security interventions often fail without practitioner input~\cite{Rauf2021}. In this RQ, we gather developer-prioritized improvements and recommendations to ensure that proposed solutions for npm package security are based on real-world practitioner experience.

\subsection{Contributions}
This study offers the following key contributions based on the real-world experiences and practices of npm package developers:
\begin{itemize}
    \item Empirical insights into how developers perceive the security of their packages and the tools and practices they use in ensuring secure development and maintenance.
    \item Insights that extend beyond purely technical barriers, uncovering human, organizational, and ecosystem-level factors that prevent effective package security.
    \item Practitioner-validated improvement priorities for secure package development and maintenance through an analysis of ranked preferences and qualitative data.
    \item The survey instrument and mixed-methods analysis provide a reusable mechanism for assessing developer security practices in other open source software ecosystems, such as PyPI, Maven, and RubyGems, enabling comparative research across communities.
\end{itemize}

\section{Background and Related Work}
\label{sec:relatedwork}

JavaScript has become a dominant language for both client- and server-side development, largely due to frameworks like Node.js and a strong developer community~\cite{bogart2016break, wittern2016look}. JavaScript projects generally fall into two categories: reusable packages integrated into other projects and standalone applications. To support the former, npm is the official package manager for Node.js, offering tools for dependency management and hosting over a million packages.

\citet{miller2024understanding} conducted a large-scale quantitative study on package abandonment, finding that it is a common phenomenon in npm. They showed that most dependents continue to rely on abandoned packages, though removal occurs more frequently when maintainers provide explicit abandonment notices. To address the challenges of analyzing such a large and dynamic ecosystem, npm-miner was developed by~\citet{chatzidimitriou2018npm}, as a platform that crawls the npm registry, assesses package quality with respect to maintainability and security, and provides a search engine to help developers evaluate packages. Decan et al.~\cite{Decan2018} conducted an empirical study on the impact of security vulnerabilities in the npm package dependency network, showing that vulnerabilities often take a long time to be discovered and fixed, with high-severity vulnerabilities taking longer to address due to their complexity. Additionally, the study highlighted the impact of vulnerable packages on dependent packages, with more than half of the dependent packages being affected by vulnerabilities in upstream packages. Liu et al. \cite {Liu2022ICSE} investigated vulnerability propagation in the NPM ecosystem. They analyzed over 50 million dependency trees and found vulnerabilities in 25\% of library versions.  

Community and contributor dynamics in NPM ecosystems have also been studied. Izquierdo and Cabot~\cite{canovas2022analysis} analyzed the 100 most popular npm projects to investigate non-coding contributions, showing that contributors often specialize in roles such as documentation or issue triage. They argue that explicit onboarding and governance strategies are essential to attract and retain these contributors, thereby improving long-term sustainability. From a supply chain perspective, Dey et al.~\cite{dey2018software} analyzed 13,000 popular npm packages and demonstrated that the download activity of both upstream and downstream runtime dependencies strongly influences package popularity. Complementing this work,~\citet{zerouali2019diversity} examined 175,000 npm packages across multiple platforms and revealed that popularity can be defined in multiple, often unrelated, ways. They highlight the need for a more comprehensive framework that combines different popularity measures.

Contribution practices have been another focus in npm ecosystems. Dey and Mockus~\cite{dey2020effect} studied 470,925 pull requests from 3,349 npm packages and identified 15 key factors influencing pull request acceptance, particularly repository-level indicators such as volume and acceptance ratios. This highlights the socio-technical aspects of pull request acceptance in npm projects. Additionally, Wittern et al.~\cite{wittern2016look} investigated dependency evolution, and report that more than 80\% of npm packages depend on at least one other package, reflecting the ecosystem's high interdependencies.

\citet{abdalkareem2017developers} studied the prevalence and risks of ``trivial'' packages—small packages implementing simple functionality. Mining over 230,000 npm packages and surveying 88 Node.js developers, they found that trivial packages make up 16.8\% of the ecosystem. While developers view them as reliable and convenient, empirical analysis revealed risks such as missing tests (fewer than half include tests) and heavy dependency chains (over 11\% depend on more than 20 other packages). 
Zimmermann et al.~\cite{zimmer} examined the impact of package dependencies, maintainer influence, and publicly reported security issues, revealing that individual packages and maintainers can significantly affect large portions of the ecosystem. They identify key threat models, including malicious packages, unmaintained legacy code, package and account takeovers, and collusion attacks. The authors also showed an increase in the attack surface and the prevalence of unpatched vulnerabilities. 
Zerouali et al.~\cite{zerouali2022impact} compared the RubyGems and npm ecosystems to quantify the impact of security vulnerabilities. They found that in both ecosystems, the number of vulnerabilities is increasing and the time required to discover them is increasing, though vulnerabilities are detected more quickly in npm than in RubyGems. Similarly, Alfadel et al.~\cite{alfadel2023empirical} examined vulnerability reporting processes in PyPI and npm and found that differences in ecosystem policies significantly influence the extent to which package vulnerabilities remain exposed.

Our study complements previous work by focusing specifically on the security dimension of the npm ecosystem, rather than package abandonment, popularity, dependency structures, or contribution practices. In addition, we examine not only developers' security perceptions but also the concrete practices, tools, and barriers they face when building and maintaining secure npm packages. Beyond confirming the fragilities of the npm ecosystem, our findings reveal how developers prioritize threats, respond to vulnerabilities, and propose ecosystem-wide improvements. Lastly, we believe our study fills a real gap, as prior research has not systematically examined npm developers’ security concerns and priorities, despite the ecosystem's central role in today's software supply chain.

\section{Method}
\label{sec:method}
In this section, we provide a detailed description of the methodology employed in our study. We used a respondent-driven sampling design to collect the data needed to address our RQs~\cite{Stol2018}. Below, we outline our survey design and validation approach, the participant recruitment process, and the analytical methods used.

\subsection{Survey Design}
We designed and hosted our online survey using Qualtrics
, configuring it to allow only one response per participant. The survey consisted of questions focused on the experiences, perceptions, practices, challenges, and recommendations of npm package developers regarding package security. In line with best practices (\cite{Kitchenham2002,kasunic2005designing,linaaker2015guidelines}), our survey questions were developed based on the objectives of our study (Section \ref{sec:goal}) and through a review of the related literature (Section \ref{sec:relatedwork}). Our survey consisted of 26 questions, including single-choice, multiple-choice, ranking, and open-ended options. Table~\ref{Table:SurveyQuestions} presents the questions included in our survey.

\begin{table*}
\centering
\caption{Survey questions included in the study. SCQ = Single-Choice Question, MCQ = Multiple-Choice Question, and an asterisk (*) indicates a mandatory question. The full questionnaire, including response options, is available at \cite{ArtifactPackage}.}
\vspace{-3mm}
\label{Table:SurveyQuestions}
\resizebox{\linewidth}{!}{%
\begin{tabular}{>{\hspace{0pt}}m{0.017\linewidth}|>{\hspace{0pt}}m{0.695\linewidth}|>{\hspace{0pt}}m{0.078\linewidth}|>{\hspace{0pt}}m{0.192\linewidth}} 
\hline
\multicolumn{1}{>{\centering\hspace{0pt}}m{0.032\linewidth}|}{\textbf{No.}} & \multicolumn{1}{>{\centering\hspace{0pt}}m{0.606\linewidth}|}{\textbf{Survey Question }(* indicates answer required)}            & \multicolumn{1}{>{\centering\hspace{0pt}}m{0.098\linewidth}|}{\textbf{Type}} & \multicolumn{1}{>{\centering\arraybackslash\hspace{0pt}}m{0.192\linewidth}}{\textbf{Notes}}  \\ 
\hline \hline
1                                                                            & Are you currently or have you ever been involved in developing, maintaining, or contributing to npm packages? *                  & Yes/No                                                                       & End survey if ``No''                                                                            \\ 
\hline
2                                                                            & Which of the following roles best describes your involvement with npm packages? *                                                & MCQ                                                                 & Includes free-text option                                                                     \\ 
\hline
3                                                                            & How long have you been actively involved with npm packages in any capacity (e.g., maintaining, contributing, etc.)? *            & SCQ                                                               &                                                                                               \\ 
\hline
4                                                                            & Approximately how many npm packages are you associated with? *                                                                   & SCQ                                                                &                                                                                               \\ 
\hline
5                                                                            & What types of npm packages are you associated with? *                                                                            & MCQ                                                                 & Includes free-text option                                                                   \\ 
\hline
6                                                                            & To the best of your knowledge, what is the approximate monthly download count across all npm packages you are associated with? * & SCQ                                                                &                                                                                               \\ 
\hline
7                                                                            & How would you rate your knowledge about JavaScript/Node.js security best practices? *                                            & SCQ                                                                &                                                                                               \\ 
\hline
8                                                                            & How important is security in your npm package development process? *                                                             & SCQ                                                                &                                                                                               \\ 
\hline
9                                                                            & What would you rate the overall security of the npm packages you maintain/contribute to? *                                       & SCQ                                                                &                                                                                               \\ 
\hline
10                                                                           & How do security concerns compare to other package development priorities (e.g., performance, functionality, usability)? *        & SCQ                                                                &                                                                                               \\ 
\hline
11                                                                           & Please rank what you believe are the most significant security threats to npm packages. (1 being the most significant) *         & Ranking                                                                      &                                                                                               \\ 
\hline
12                                                                           & What, if any, are other areas you perceive as significant threats to npm package security?                                       & Free-Text                                                                    &                                                                                               \\ 
\hline
13                                                                           & How satisfied are you with the current available tools to identify and remove security vulnerabilities in npm packages? *        & SCQ                                                                &                                                                                               \\ 
\hline
14                                                                           & Please explain why you selected the above answer concerning security tools. *                                                    & Free-Text                                                                    &                                                                                               \\ 
\hline
15                                                                           & Which of the following security practices do you follow for your npm packages? *                                                 & MCQ                                                                 & Includes free-text option                                                                     \\ 
\hline
16                                                                           & Which of the following tools do you use to identify security vulnerabilities in your package(s)? *                               & MCQ                                                                 & Includes free-text option                                                                     \\ 
\hline
17                                                                           & How often do you run the selected tool(s)? *                                                                                     & SCQ                                                                & Shown if \textit{not} ``None'' (Q16)\par{}Includes free-text option                        \\ 
\hline
18                                                                           & How often are the dependencies in your npm packages updated or audited for security vulnerabilities? *                           & SCQ                                                                & Includes free-text option                                                                     \\ 
\hline
19                                                                           & Have you ever stopped using a dependency due to security concerns? *                                                             & Yes/No                                                                       &                                                                                               \\ 
\hline
20                                                                           & What was the reason(s)? *                                                                                                        & MCQ                                                                 & Shown if ``Yes'' (Q19)\par{} Includes free-text option                            \\ 
\hline
21                                                                           & How do/would you typically respond when a security vulnerability is discovered in your npm package? *                                  & MCQ                                                                 & Includes free-text option                                                                     \\ 
\hline
22                                                                           & What are the main challenges you face in ensuring the security of your npm packages? *                                           & MCQ                                                                 & Includes free-text option                                                                    \\ 
\hline
23                                                                           & What challenges do you face with existing security tools? *                                                                      & MCQ                                                                 & Includes free-text option                                                                    \\ 
\hline
24                                                                           & Please rank the following areas based on how impactful they would be in improving npm package security. (1 being the most) *     & Ranking                                                                      &                                                                                               \\ 
\hline
25                                                                           & What, if any, are other impactful areas that would improve npm package security?                                                 & Free-Text                                                                    &                                                                                               \\ 
\hline
26                                                                           & Based on your experience, what security practices, if any, would you recommend to other npm package maintainers/contributors?    & Free-Text                                                                    &                                                                                               \\
\hline \hline
\end{tabular}
}
\end{table*}

\subsection{Pilot Run}
Before publicly launching our survey, we conducted a pilot run to assess its validity~\cite{MOLLERI2020}. During the pilot run, we enlisted five developers from the author's professional network. The participants were asked to evaluate the clarity of the questions and the overall flow of the survey and to provide unbiased, constructive feedback. This process allowed us to identify the shortcomings in our questionnaire, particularly the questions that needed rewording or removal, to improve its clarity. For instance, 
we transformed question \#24 from multiple-choice into ranking questions based on the input we received. It's important to note that the responses from the pilot run were discarded before launching the final survey, ensuring our analysis relied solely on data from actual respondents.

\subsection{Sampling Strategy and Recruitment}

For this study, we used a purposive sampling approach~\cite{Baltes2022}, specifically selecting developers who publish and maintain packages in the npm registry. We obtained their contact information using the official npm registry API~\footnote{\url{https://github.com/npm/registry}}. This method ensures that respondents are directly involved in package development, making them suitable for our study goals. However, since not all npm developers publicly share their contact details, our sample may not fully represent the entire npm developer community. Participation in the survey was voluntary and anonymous, with no compensation provided. To achieve a pragmatic balance between coverage and feasibility, we sent personalized email invitations to 1,000 npm package developers between late Spring 2025 to early Fall 2025 in small batches\footnote{We had to adhere to our institution's email constraints, which limited the frequency and amount of bulk emails sent to external recipients.}. In total, we received 130 responses.
To maintain consistency in our analysis, we excluded four respondents who failed the prescreening question (answered ``No'' to survey question \#1). We then included only those who completed all mandatory questions, thereby improving the reliability and validity of our findings. After excluding incomplete responses, we had \SurveyParticipantCount usable responses for our RQs.

\subsection{Data Analysis}

We employed a mixed-methods approach that combined quantitative and qualitative techniques to analyze the survey data~\cite{Wagner2020}. Quantitative analysis included descriptive statistics and frequency analysis for single-choice and multiple-choice questions. We did not perform statistical imputation. The Borda Count method was used to rank preferences in questions with ordered responses, assigning points based on each option's rank and calculating proportions relative to the maximum possible score~\cite{Saari2022}. For qualitative analysis, we manually reviewed free-text responses from survey respondents to identify patterns and themes. This process followed a structured approach, and two of the authors, both experienced software engineering researchers, independently analyzed and coded the responses to ensure reliability. They documented the emerging themes and resolved any differences in interpretation through discussion.  Instead of calculating a quantitative measure of inter-rater reliability, robustness was achieved through discussion. This negotiated agreement approach is recognized as effective for ensuring coding quality in qualitative research~\cite{Armstrong1997, Campbell2013}. 

\section{Results}
\label{sec:results}

In this section, we report our results~\footnote{Due to space constraints, we only present the most common findings in certain areas of the write-up. The complete breakdown is in our dataset at \cite{ArtifactPackage}.}. Before discussing our RQs, we outline the experience and qualifications of the \SurveyParticipantCount survey respondents based on their answers to survey questions \#1 to \#6.

All \SurveyParticipantCount respondents answered ``Yes'' to question \#1, indicating that they have experience in npm package development. In response to question \#2, 70 respondents (54.26\%) identified themselves as maintainers, while 52 (40.31\%) identified as contributors. Additionally, five respondents indicated that they were security specialists. Since this was a multiple-choice question, the median number of roles chosen was two, with the maintainer and contributor roles frequently selected together (48 occurrences).

For question \#3, 35 respondents (46.67\%) reported having 7 to 10 years of experience in npm package maintenance, followed by 18 respondents with 4 to 6 years, 15 with 1 to 3 years, and seven with less than one year. From question \#4, 30 respondents (40\%) are associated with 2 to 5 packages, 21 (28\%) with more than 20 packages, 13 with 6 to 10 packages, five with 11 to 20, and six with just one package. In question \#5, the most common types of packages are utility libraries (50 responses; 22.42\%), front-end libraries/frameworks (46 or 20.63\%), and back-end libraries/frameworks (39 or 17.49\%). As this was a multiple-choice question, respondents selected a median of 3 options, with 19 choosing the three aforementioned types. 
Lastly, the monthly download counts of the respondents' packages are shown in Table \ref{Table:monthlyDownloads}. Among the respondents, 18 (24\%) reported fewer than 1,000 downloads. Additionally, 13 respondents reported having between 10,001 and 100,000 downloads, while 12 respondents reported having between 1,000,001 and 10,000,000 downloads.

\begin{table}
\centering
\caption{Monthly downloads of respondents' packages.}
\vspace{-1mm}
\label{Table:monthlyDownloads}
\resizebox{\columnwidth}{!}{%
\begin{tabular}{@{}p{0.65\linewidth}rr@{}} 
\toprule
\multicolumn{1}{c}{\textbf{Monthly Downloads}} & \multicolumn{1}{c}{\textbf{Count}} & \multicolumn{1}{c}{\textbf{Percentage}} \\ \midrule \hline
Less than 1,000         & 18 & 24\%  \\ \midrule
1,000 - 10,000          & 12 & 16\% \\ \midrule
10,001 - 100,000        & 13 & 17.33\%  \\ \midrule
100,001 - 1,000,000     & 11 & 14.67\%  \\ \midrule
1,000,001 - 10,000,000  & 12 & 16\%  \\  \midrule 
More than 10,000,000    & 9  & 12\%  \\  \midrule \hline
\end{tabular}%
}
\end{table}

The data presented above indicates that the respondents in our survey are experienced npm package developers, increasing the likelihood of capturing real-world experiences, practices, and challenges related to securing packages within the npm ecosystem.

\subsection{\RQA}   
This RQ explores how npm developers perceive and prioritize package security. We answer this RQ through four specialized areas.

\subsubsection{\textbf{Developers’ Self-Assessment of Security Expertise and Package Security.}} 
We address this area by analyzing responses to survey questions \#7 to \#10 to understand how developers assess their security expertise and the perceived security of their packages.

Most of the respondents perceived themselves as experienced in the Node/npm context, with ``Advanced'' being the most common level of expertise in 33 of \SurveyParticipantCount responses (44\%), followed by ``Intermediate'' (24 responses; 32\%). Only two respondents identified as ``Novice,'' indicating a generally confident group of respondents. Consistent with this self-perception, respondents overwhelmingly viewed security as an important quality attribute (question \#8); 77.33\% rated it as ``Important'' or ``Extremely Important'', while only two respondents viewed it as ``Not a priority.''

However, this confidence does not fully translate into perceptions of their own packages' robustness. When asked to rate their package's overall security (question \#9), the most common response was ``Somewhat Secure'' (31 responses; 41.33\%), with only 24 respondents (32\%) describing their packages as ``Very Secure.'' Finally, when comparing security with other quality attributes like performance or usability (question \#10), developers prioritized security, with 28 respondents ranking it as ``High Priority'' or ``Essential.''  

Overall, these findings suggest that while developers are aware of security and highly value security, they recognize the gaps between the desired and actual security levels in their own packages.

\subsubsection{\textbf{Developers’ Perceptions of the Most Concerning Security Threats in the npm Ecosystem.}}  
For this area, we analyze responses to survey question \#11, which asked respondents to rank eight answer options related to different security threats that could affect a package.  To determine the overall ranking, we use the Borda Count method \cite{Saari2022}, which has been utilized in prior software engineering research on ranked survey data \cite{Peruma2024Assert}.

Table~\ref{Table:rq1b_borda} displays the significant threats ranked along with their corresponding scores. The Borda score column shows the total points each option received based on respondents' ratings. The column ``Prop.'' (proportion) compares each option's score to the maximum possible score of 2,700 (calculated as the number of respondents, \SurveyParticipantCount, multiplied by 8 points for a first-place ranking).  

\begin{table}
\centering
\caption{Ranked security threats to npm packages.}
\vspace{-1mm}
\label{Table:rq1b_borda}
\normalsize
\begin{tabular}{@{}rlrr@{}}
\toprule
\multicolumn{1}{c}{\textbf{Rank}}&\multicolumn{1}{c}{\textbf{Answer Option}} & \multicolumn{1}{c}{\textbf{Borda Score}} & \multicolumn{1}{c}{\textbf{Prop.}} \\ \midrule \hline
1 & Supply chain attacks & 467 & 0.78 \\ \midrule
2 & Dependency vulnerabilities           & 461 & 0.77 \\ \midrule
3 & Malicious code injection             & 373 & 0.62 \\ \midrule
4 & Account takeovers           & 342  & 0.57 \\ \midrule
5 &Outdated dependencies & 301  & 0.50 \\ \midrule
6 &Insufficient code review & 275  & 0.46 \\ \midrule
7 &Insufficient security testing & 239  & 0.40 \\ \midrule
7 &Package name typosquatting & 239  & 0.40 \\ \midrule \hline
\end{tabular}%
\end{table}

The findings revealed that supply chain attacks ranked as the main concern, followed closely by dependency vulnerabilities and malicious code injection, which are ranked second and third, respectively. In addition, the two top-ranked threats received very similar scores, indicating a high level of concern among developers.

Next, respondents were presented with an optional free-text response question (\#12) to specify other areas they perceive as significant security threats to npm packages. We received 12 responses to this question, and a thematic analysis of their responses revealed concerns related to both technical vulnerabilities and human factors. The themes identified are:
\begin{itemize}
    \item \textbf{Developer Security Knowledge Deficit.} Gaps in security education can lead to unsafe practices (e.g., ``\textit{many people are doing an NPM install with sudo}'') and a lack of awareness of security best practices and concepts, such as those outlined by OWASP. 
    \item \textbf{Supply Chain Infrastructure Vulnerabilities.} Respondents expressed concerns about the security risks associated with trusting multiple package registries and mirrors. While decentralized package delivery systems offer flexibility and accessibility, they also increase the potential attack surface, as a compromised mirror could deliver malicious packages. Specific vulnerabilities mentioned include spoofing attacks, cache poisoning, and registry takeover. Further, there were concerns about the disconnect between source code repositories and the published packages.
    \item \textbf{Security Tool Issues.} The primary concern is alert fatigue caused by ``\textit{too much noise}'' in security notifications, where the volume of alerts can make it difficult to identify and prioritize genuine security threats.
    \item \textbf{Ecosystem Fragmentation.} Ensuring support for multiple package managers (e.g., npm, pnpm, yarn) and JavaScript runtimes (e.g., Node, Deno, Bun) can make package maintenance challenging, which can distract developers from security concerns.  
    \item \textbf{Malicious Code Execution.} One primary concern raised by respondents is the automatic execution of post-install scripts by external packages, which can run with the same privileges as the installation process, potentially acting as a vector for arbitrary code execution without explicit user consent or review.
    \item \textbf{Package Maintainer Trust.} Respondents expressed concerns regarding package maintainers, including those who intentionally insert malicious code, as well as those whose behavior might not be malicious but still creates security risks (e.g., ``\textit{drunk, or crazy package maintainers}''). 
    \item \textbf{Dependency Freshness Challenges.} Widely used but unmaintained packages pose security risks due to unpatched vulnerabilities. While dependency version pinning can mitigate this, it may lead to compatibility issues and technical debt (e.g., ``\textit{Unmaintained software causing pinning of dependencies}'').

\end{itemize}

\subsubsection{\textbf{Developers’ Satisfaction with the Current Security Tools.}}  
We address this area through survey questions \#13 and \#14, which examine the extent to which developers are satisfied with current security tools.

\begin{figure}[!ht]
    \centering
    \includegraphics[width=1\linewidth]{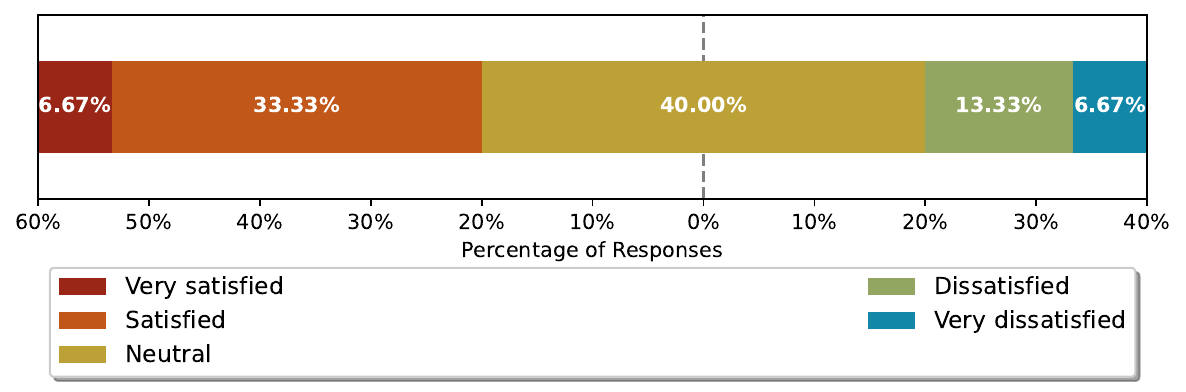}
    \caption{Satisfaction with security tools for npm packages.}
    \label{Figure:rq1c_likert}
\end{figure}

In Figure~\ref{Figure:rq1c_likert}, it is concerning to observe that only 30 (or 40\%) respondents are either ``Satisfied'' or ``Very Satisfied'' with the current set of tools to identify and remove security vulnerabilities in npm packages. This distribution indicates that while most developers acknowledge the value of existing tools, overall satisfaction remains lukewarm, and the prevalence of ``Neutral'' and ``(Very) Dissatisfied'' responses highlights challenges with current security tooling.

Next, respondents were asked to justify their satisfaction response via a free-text response question (\#14).

First, we analyzed the free-text responses from respondents who were either ``Very dissatisfied'' or ``Dissatisfied'' with the current set of security tools. From the 15 responses, one was discarded due to being unclear. Some of the key reasons for dissatisfaction include:
\begin{itemize}
    \item \textbf{Tool Noise and Alert Fatigue.} Respondents complained that security scanners generate too many false positives or contextually irrelevant warnings, as one respondent noted, ``\textit{The security scanning system in npm is a complete joke and more of a nuisance than anything. 99\% of the “vulnerabilities” are idiotic and not worthy of a real CVE}.'' These excessive or low-value alerts may be counterproductive, as they divert developers from conducting meaningful work. As another remarked,  ``\textit{way to noising and causing lots of work for maintainers}.''
    \item \textbf{Feature Gaps and Usability Limitations.} Respondents noted that existing tools are largely reactive and fail to detect common coding mistakes or patterns that lead to vulnerabilities. For instance, one participant highlighted the absence of ``\textit{common mistake scanning (pattern matching, etc.)},'' while another criticized that ``\textit{tooling feels primitive and buggy}.'' 
    \item \textbf{Limited Tool Awareness.} Some respondents admitted that they lacked knowledge about available tools, with responses like ``\textit{Do not know any good tools}'' and ``\textit{I don't use it}.''
\end{itemize}

Next, when examining responses from respondents who rated their satisfaction as ``Neutral,'' we found that the reasons were quite similar to those given by dissatisfied respondents. After removing three unclear responses, we found that ten out of the remaining 27 responses were associated with Limited Tool Awareness, six were related to Feature Gaps and Usability Limitations, and four were linked to Tool Noise and Alert Fatigue.

Finally, we looked at the responses from those who rated their satisfaction as ``Very Satisfied'' or ``Satisfied.'' After removing four unclear responses, we identified 15 replies where Tool Effectiveness was cited as the reason for their satisfaction. For example, one noted, ``\textit{Socket and Snyk are doing a great job},'' and another mentioned, ``\textit{existing tools work fairly well}.'' We also noted four responses related to Feature Gaps and Usability Limitations, where even satisfied users highlighted areas for improvement, such as ``\textit{npm audit is effective at enumerating and automatically fixing vulnerabilities, but lacks realtime scanning and alerts}.'' Additionally, we encountered responses categorized under Human Factors in Security, focusing on the developer or maintainer, with remarks like ``\textit{vlnerabilities in code aren't the cause of any significant npm incidents - 100\% of them involved either a rogue/burnt out maintainer, or an account takeover}'' and ``\textit{people are always the weak link!}''

\vspace{2mm}
\noindent\textbf{\textit{Summary for RQ1.}}
Developers viewed security as important and essential, yet most rate their own packages only ``Somewhat Secure.'' The primary security concerns include supply chain attacks, dependency vulnerabilities, and malicious code injection. Only 40\% of developers are satisfied with the current security tools for npm packages. Common issues include alert fatigue, feature gaps, and a lack of awareness about available tools.

\subsection{\RQB}
Having established developers' security perceptions, this RQ examines their actual security practices and tool usage in development workflows, including their typical responses when vulnerabilities are identified. We address this RQ through three areas.

\subsubsection{\textbf{Commonly Adopted Security Practices and Tools Among npm Developers.}}  

In this area, we first analyze the security practices that developers follow (question \#15). As shown in Table \ref{Table:rq2_securityPractices}, the most prevalent practices were automated methods, such as two-factor authentication for npm accounts, followed by automated vulnerability scanning and security linting. However, human-oriented practices, including security-related code reviews and security trainings, were less frequently adopted. This discrepancy highlights areas that require intervention, particularly in integrating human elements into security practices. As a multiple-choice question, respondents selected, on average, three practices, with 30 choosing the three aforementioned types. In the ``Other'' category, one respondent noted using AI for reviews, and another mentioned reducing dependencies.

Next, from question \#16, the most commonly used tools are npm audit (53 responses; 34.19\%) and Dependabot (50; 32.26\%). Snyk appears in about a quarter of responses (19; 12.26\%), with a smaller adoption of SonarQube and OWASP Dependency-Check (each 9; 5.81\%). In particular, eight respondents reported that they do not use any security-related tools. In the ``Other'' category, the respondents mentioned Spice Labs (1 response) and Socket (2 responses). On average, developers utilize two tools: 38 reported using both Dependabot and npm audit, while 15 reported using Snyk in combination with npm audit. For tool frequency (question \#17), most run them automatically using Dependabot or a similar tool (29 responses; 38.67\%), followed by daily (11; 14.67\%) and weekly (10; 13.33\%). Only seven reported running tools only when issues arise.

\begin{table}
\centering
\caption{Top five security practices followed by developers.}
\vspace{-1mm}
\label{Table:rq2_securityPractices}
\resizebox{\columnwidth}{!}{%
\begin{tabular}{@{}p{0.61\linewidth}rr@{}} 
\toprule
\multicolumn{1}{c}{\textbf{Individual Answer Option}} & \multicolumn{1}{c}{\textbf{Count}} & \multicolumn{1}{c}{\textbf{Percentage}} \\ \midrule \hline
Two-factor authentication for npm account         & 60 & 27.03\%  \\ \midrule
Automated vulnerability scanning & 49  & 22.07\% \\ \midrule
Security linting   & 43  & 19.37\%  \\ \midrule
Code reviews focused on security before publishing                & 29  & 13.06\%  \\ \midrule
Security training for team members                   & 18  & 8.11\%  \\  \midrule \hline
\end{tabular}%
}
\end{table}

\subsubsection{\textbf{Security-Driven Dependency Decisions.}}  
In this area, we examine dependency hygiene cadence and the rationale for dependency discontinuation.

For dependency updates/audits (question \#18), the most common response was ``Monthly'' (15 responses; 20\%). Remarkably, ``Never'' was the second most popular choice, receiving 14 responses. The ``Other'' category also had 14 responses, with respondents describing their approaches as event- or release-driven. Examples included responses like ``\textit{when flagged by Dependabot or noticed by users or developers},'' ``\textit{on every pull request},'' ``\textit{whenever we receive a feature request},'' or varying based on the project.

Next, when asked if they had ever stopped using a dependency due to security concerns (question \#19), 51 (68\%) responses were ``Yes''. These 51 respondents were subsequently asked to elaborate on the reasons behind their decision to discontinue the use of the dependency (question \#20). As shown in Table~\ref{Table:rq3_stopDependency}, 41 (40.20\%) cited package abandonment or deprecation. This was followed by issues related to packages with unpatched vulnerabilities and non-responsive maintainers. Additionally, the responses reported under ``Other'' highlighted maintainer issues, such as ``\textit{not well maintained}'' and ``\textit{maintainers sabotaging their own package}.''

\begin{table}
\centering
\caption{Reason for stopping the use of a dependency.}
\vspace{-1mm}
\label{Table:rq3_stopDependency}
\resizebox{\columnwidth}{!}{%
\begin{tabular}{@{}p{0.61\linewidth}rr@{}} 
\toprule
\multicolumn{1}{c}{\textbf{Individual Answer Option}} & \multicolumn{1}{c}{\textbf{Count}} & \multicolumn{1}{c}{\textbf{Percentage}} \\ \midrule \hline
Package abandoned or declared deprecated & 41  & 40.20\% \\ \midrule
Unpatched vulnerabilities         & 29 & 28.43\%  \\ \midrule
Maintainer unresponsive to security issues  & 28  & 27.45\%  \\ \midrule
Other                 & 4  & 3.92\% \\  \midrule \hline
\end{tabular}%
}
\end{table}

\subsubsection{\textbf{Responses to Security Vulnerabilities.}}  
The final analysis for this RQ examines how developers respond to vulnerabilities in their packages, which we assess through survey question \#21.

Table~\ref{Table:rq2_response} shows that respondents usually assessed the severity before taking action (26.67\%) and then fixed and released a patch immediately (23.08\%).  Less frequent actions include publishing advisories (7.18\%), privately notifying affected users before a fix (4.10\%), and temporarily unpublishing the package (1.54\%).

\begin{table}
\centering
\caption{Top five responses to security vulnerabilities.}
\vspace{-1mm}
\label{Table:rq2_response}
\resizebox{\columnwidth}{!}{%
\begin{tabular}{@{}p{0.58\linewidth}rr@{}} 
\toprule
\multicolumn{1}{c}{\textbf{Individual Answer Option}} & \multicolumn{1}{c}{\textbf{Count}} & \multicolumn{1}{c}{\textbf{Percentage}} \\ \midrule \hline
Assess the severity of the vulnerability before taking action         & 52 & 26.67\%  \\ \midrule
Fix and release patch immediately & 45  & 23.08\% \\ \midrule
Document the fix in release notes   & 32  & 16.41\%  \\ \midrule
Audit dependencies for transitive vulnerabilities                & 21 & 10.77\%  \\ \midrule
Delay fixing due to time/resource constraints                 & 15  & 7.69\%  \\  \midrule \hline
\end{tabular}%
}
\end{table}

\noindent\textbf{\textit{Summary for RQ2.}}
Developers commonly adopt automated security practices, such as two-factor authentication and automated vulnerability scanning, more than human-oriented practices, such as security-focused code reviews and security training. Tools such as npm audit and Dependabot are widely used, with many developers running them automatically, while some reported not using any security tools at all. Package abandonment and unpatched vulnerabilities are the main reasons for discontinuing dependencies. When vulnerabilities arise, developers typically assess severity and quickly release patches.

\subsection{\RQC}
In this RQ, we aim to understand the general and tool-specific barriers that npm developers face when securing their packages, as captured in survey questions \#22 and \#23.

As shown in Table~\ref{Table:rq3_challenges}, respondents most frequently cited time constraints as a key barrier (49 responses; 26.2\%). Other notable challenges included difficulty keeping up with security updates and emerging threats (33; 17.6\%) and the complexity of managing dependencies (23; 12.3\%). Insufficient community support was the least cited issue, reported by only 11 respondents. In addition, security testing and balance of security with other quality attributes each received 14 responses. On average, respondents selected approximately three distinct challenges, underscoring that obstacles are multifaceted rather than isolated. 

\begin{table}
\centering
\caption{Top five developer challenges in securing packages.}
\vspace{-1mm}
\label{Table:rq3_challenges}
\resizebox{\columnwidth}{!}{%
\begin{tabular}{@{}p{0.64\linewidth}rr@{}} 
\toprule
\multicolumn{1}{c}{\textbf{Individual Answer Option}} & \multicolumn{1}{c}{\textbf{Count}} & \multicolumn{1}{c}{\textbf{Percentage}} \\ \midrule \hline
Time constraints         & 49 & 26.20\%  \\ \midrule
Difficulty keeping up with security updates and threats & 33  & 17.65\% \\ \midrule
Complexity of managing dependencies   & 23  & 12.30\%  \\ \midrule
Lack of awareness or understanding of security best practices                & 19  & 10.16\%  \\ \midrule
Limited access to security resources                   & 17  & 9.09\% \\  \midrule \hline
\end{tabular}%
}
\end{table}

At the tool level (Table \ref{Table:rq3_challengesTools}), the obstacle most frequently reported was a high false-positive rate in security scans (35 responses; 30.2\%), followed by inaccurate vulnerability detection and limited automation for dependency management. The least reported issue was integration difficulties with CI/CD pipelines (7 responses). Comments in the ``Other'' category included licensing constraints, overreliance on dependencies, and tools limited to static analysis.

\begin{table}
\centering
\caption{Top five developer challenges with security tools.}
\vspace{-1mm}
\label{Table:rq3_challengesTools}
\resizebox{\columnwidth}{!}{%
\begin{tabular}{@{}p{0.61\linewidth}rr@{}} 
\toprule
\multicolumn{1}{c}{\textbf{Individual Answer Option}} & \multicolumn{1}{c}{\textbf{Count}} & \multicolumn{1}{c}{\textbf{Percentage}} \\ \midrule \hline
High false positive rate in security scans & 35  & 30.17\% \\ \midrule
Lack of accurate vulnerability detection         & 25 & 21.55\%  \\ \midrule
Limited automation for dependency management and updates  & 20  & 17.24\%  \\ \midrule
Lack of actionable guidance for fixing vulnerabilities              & 19  & 16.38\%  \\ \midrule
Other                 & 10  & 8.62\% \\  \midrule \hline
\end{tabular}%
}
\end{table}

\vspace{2mm}
\noindent\textbf{\textit{Summary for RQ3.}}
Time constraints are the most frequently cited barrier to secure package development, with other challenges including difficulties in keeping up with security updates and managing dependencies. At the tool level, a high false-positive rate in security scans was the most frequently reported issue, while CI/CD integration issues are comparatively rare. 

\subsection{\RQD}
While the prior RQs focused on current perceptions, practices, and challenges in secure package development, this RQ aims to identify improvements that developers believe would further strengthen package security. We address it through survey questions \#24-\#26.

Starting with survey question \#24, which asks respondents to rank areas for improvement, we see from Table \ref{Table:rq4_borda} that the top priority is better automated vulnerability detection tools. This is followed by the need for comprehensive, up-to-date security documentation and best practices, as well as financial support or incentives to help maintain secure packages. The lowest priority, on the other hand, is tools that automatically generate fixes for vulnerabilities.

The substantial gap between the first and last items (a 226-point Borda difference and a 0.43 proportion gap) highlights that developers strongly value accurate, reliable detection tools but are less enthusiastic about automated fixing. One possible interpretation is that developers lack trust in automatically generated fixes. This suggests a cautious approach to handling security vulnerabilities and a preference for manual interventions to ensure that fixes are appropriate, context-aware, and do not inadvertently introduce new issues. Additionally, the emphasis on financial and documentation-related priorities highlights that maintaining security is not solely a tooling challenge but also a resourcing and knowledge problem.    

\begin{table}
\centering
\caption{Ranked improvement priorities for package security.}
\vspace{-1mm}
\label{Table:rq4_borda}
\resizebox{\columnwidth}{!}{%
\normalsize
\begin{tabular}{@{}rp{0.53\linewidth}rr@{}}
\toprule
\multicolumn{1}{c}{\textbf{Rank}}&\multicolumn{1}{c}{\textbf{Answer Option}} & \multicolumn{1}{c}{\textbf{Borda Score}} & \multicolumn{1}{c}{\textbf{Prop.}} \\ \midrule \hline
1 & Improved automated vulnerability detection tools & 434 & 0.83 \\ \midrule
2 & Comprehensive, up-to-date security documentation and best practices           & 372 & 0.71 \\ \midrule
3 & Financial support or incentives for maintaining secure packages             & 303 & 0.58 \\ \midrule
4 & Better vulnerability reporting systems           & 272  & 0.52 \\ \midrule
5 & Improved dependency management & 256  & 0.49 \\ \midrule
6 & Specialized developer security training & 255  & 0.46 \\ \midrule
7 &Tools that automatically generate fixes for vulnerabilities & 208  & 0.40 \\ \midrule \hline
\end{tabular}%
}
\end{table}
 
Next, through an optional free-text question (\#25), respondents were invited to suggest additional impactful areas that could enhance npm package security, resulting in six responses. The analysis of these responses highlighted several key areas for improvement. Respondents noted the importance of mandatory account hardening for maintainers by implementing two-factor authentication. They also suggested improved vulnerability reporting through multiple channels and a community-driven system. Additionally, respondents recommended expanding the standard/core Node.js libraries to include functionalities found in third-party dependencies, which would help reduce the overall number of dependencies. Limiting the publishing of packages to trustworthy package managers and implementing human-in-the-loop reviews for popular packages were also mentioned as vital improvements.

Finally, through another optional free-text question (\#26), respondents were invited to share their security recommendations with the community. We received 15 responses to this question, and a thematic analysis yields the following recommendations:
\begin{itemize}
    \item \textbf{Secure Maintainer Account Practices.} There was a strong emphasis on account security, with advice ranging from enabling two-factor authentication to preventing phishing attacks. This indicates a growing awareness of the risks associated with account hijacking as a significant attack vector. Additionally, recommendations included establishing governance around account access and implementing restrictions on package publishing.
    \item \textbf{Dependency Management and Reduction.} Supply chain complexity is another prevalent security concern. Several respondents recommended minimizing the number of dependent packages, pinning versions, carefully evaluating complex dependencies, and examining transitive dependencies.
    \item \textbf{Tooling and Automation.} The use of tools for continuous scanning of vulnerabilities is another recommendation from respondents. A respondent also recommended further research on automated security tools, noting that the current set is insufficient for comprehensive protection against vulnerabilities.
    \item \textbf{Code Reviews.} There was also a recommendation for conducting continuous code reviews and thorough contributor vetting.
    \item \textbf{Security Awareness and Education.} Respondents highlighted the necessity for ongoing and in-depth security knowledge. This includes adopting a practical and offensive mindset to better defend against vulnerabilities. As one respondent noted, ``\textit{If you know how to do a SQL injection attack for example, you probably aren't going to write a SQL injection vulnerability}.''
    \item \textbf{Incentive-Based Security Culture.} There was a recommendation to promote a security-oriented culture through bug bounty programs and financial incentives, highlighting that effective security requires commitment.
\end{itemize}

\vspace{2mm}
\noindent\textbf{\textit{Summary for RQ4.}}
Developers prioritize better automated vulnerability detection tools over other improvements. In contrast, there is less enthusiasm for automated fixing tools. Additional suggestions to improve package security include mandatory two-factor authentication for maintainer accounts and improved vulnerability reporting systems, among others. Developers also emphasized the importance of security awareness, education, and incentives/bug bounties to reinforce a security culture.

\section{Discussion}
\label{sec:discussion}

Our findings shed light on the perceptions and challenges npm developers face regarding package security. This section discusses the implications of these results, connects them to prior research, and highlights opportunities for improvement.

\noindent\textbf{The Awareness–Practice Gap in npm Package Security.} Across all RQs, our findings show a consistent intention–action gap in how npm developers approach package security. Most respondents rate security as important, aligning with prior evidence that practitioners recognize the value of secure software development~\cite{torten2018impact}. However, they rate their own packages as only ``somewhat secure.'' This mismatch highlights that awareness alone is insufficient to ensure secure practices and may stem from structural and temporal constraints, such as limited time and high workloads, that hinder security efforts. Further, the complexity of the npm ecosystem complicates these challenges, indicating that npm package security involves not just technical issues, but also human, organizational, and ecosystem factors that require a more comprehensive approach.

\noindent\textbf{Supply Chain Vulnerabilities and Ecosystem Fragility.} Our findings confirm that supply chain attacks and dependency vulnerabilities are developers’ primary concerns in the npm ecosystem, with free-text responses describing trust issues with maintainers, unmaintained dependencies, and risky post-install scripts. These results align with broader research showing that attackers exploit three main vectors: injecting vulnerabilities into dependencies, compromising build infrastructure, and targeting developers through social engineering~\cite{williams2025research}. 
Addressing these issues necessitates better auditing tools, registry monitoring, and enhanced governance and community practices for secure maintenance.

\noindent\textbf{Dependency Problems and Discontinuation Decisions.}
There is notable variability in dependency update practices among developers: some adopt proactive, automated strategies, while others never update dependencies unless prompted by external events. This variability increases systemic risk, as outdated and abandoned dependencies persist in the ecosystem. When developers do discontinue dependencies, the most frequent drivers are package abandonment and unpatched vulnerabilities, further highlighting the fragility of the dependency network. Prior research shows that npm has single points of failure, where a few maintainers can affect many packages, and vulnerable, unmaintained dependencies persist for years~\cite{zimmer}. These findings reinforce our respondents’ concerns that dependency fragility and maintainer trust are critical risk factors, highlighting the need for stronger mechanisms to signal package health and maintenance status, reduce reliance on abandoned dependencies, and mitigate single-maintainer risks.

\noindent\textbf{Adoption of Security Tools.} Tool usage in the ecosystem is uneven, with basic tools like npm audit and Dependabot being more popular than advanced third-party options. Developers often express dissatisfaction due to issues such as false positives, insufficient actionable guidance, and alert fatigue. Although automation is appreciated, concerns about usability and trust hinder wider adoption. This is especially critical in the npm ecosystem, where vulnerabilities arise from risks such as malicious install scripts, expired maintainer accounts, and inactive maintainers that leave packages vulnerable to takeover \cite{Zahan}. The findings highlight the need for more precise, context-aware security tools that not only reduce noise but also surface systemic risks, thereby strengthening developer confidence in security interventions.

\noindent\textbf{Socio-Technical Security Debt.} Our findings consistently demonstrate that purely technical security measures are weakened by human and organizational factors, creating what we term \textit{socio-technical security debt}. In the npm ecosystem, this debt builds up in several ways. For example, alert fatigue from false positives leads to a gradual decline in the effectiveness of security tools, as shown by developers’ dissatisfaction with these tools. Time constraints often force developers to postpone important security tasks. Further, gaps in security knowledge and best-practice guidance mean that even well-intentioned developers might accidentally introduce or overlook risks. Together, these factors interlace technical, human, and organizational weaknesses into a cycle of debt that undermines the ecosystem's overall security.

\noindent\textbf{Human–AI Collaboration.} Our findings highlight the need for precise, context-aware automation. While large language models (LLMs) are being used in software development, they exhibit limited security awareness and vulnerability reasoning \cite{sajadi2025llms}. Future research should explore hybrid human–AI security workflows to enhance developer judgment and balance automation with human oversight, helping to reduce socio-technical security debt.  

\noindent\textbf{Beyond the NPM Ecosystem.} While our findings focus on npm, challenges like time constraints, dependency complexity, and frustration with security tools are likely common across other OSS package ecosystems (e.g., PyPI, Maven, RubyGems). Recognizing these common issues can lead to improved tools and practices for better developer efficiency and security across platforms.

\noindent\textbf{Practical Implications for Stakeholders} 
\begin{itemize}
    \item \textbf{Maintainers and Contributors.} Security must be proactive and integrated into routine workflows. Dissatisfaction with security tools suggests that maintainers must actively tune them to reduce false positives or expand their security toolset. Developers should supplement automation with simple manual checks, like peer code reviews, dependency hygiene, and issue triage to avoid minor vulnerabilities turning into bigger problems. Further, to encourage new contributions, maintainers should label security tasks as ``good first issues.''
    Establishing governance practices, such as pre-release checklists and contributor guidelines, can promote shared responsibility among contributors.
    \item \textbf{Ecosystem Stewards.} Registry providers such as npm and GitHub can reduce systemic risk by enforcing two-factor authentication for maintainers, improving multi-channel vulnerability reporting, and displaying package health and maintainer activity indicators. Enhanced documentation and best-practice repositories would also support maintainers in adopting secure behaviors. Additionally, financial support such as grants, recognition programs, or bug bounty incentives can help address security debt. 
    \item \textbf{Tool Builders.} Security tool builders should prioritize precision and usability to minimize alert fatigue and promote developer trust. These tool builders should focus on designing for human-AI collaboration rather than full automation, providing contextual guidance, severity calibration, and actionable remediation, rather than limiting themselves to presenting lists of vulnerabilities.
    \item \textbf{Regulators and Policy Makers.} Emerging cybersecurity regulations, such as the EU Cyber Resilience Act \cite{EUcyber}, define standards for vulnerability disclosure and secure OSS development. However, our findings show that maintainer capacity and tool usability may not meet these standards, highlighting the need for practical guidance and support for volunteer-driven projects.
    \item \textbf{Researchers.} The socio-technical security debt concept offers research opportunities that extend beyond traditional security metrics. These opportunities include frameworks for measuring this debt, longitudinal studies to track accumulation over time, and intervention strategies. Additionally, comparative studies across ecosystems may reveal effective governance structures and practices to minimize the accumulation of this debt.
    \item \textbf{Educators.} The gap between self-assessed expertise and actual security outcomes indicates that many package developers may have incomplete mental models of security threats. It is essential to address these gaps in security fundamentals to ensure that developers' security awareness translates into secure practices. 
\end{itemize}

\vspace{-5mm}
\section{Threats to Validity}
\label{sec:limitations}

Below, we describe some key threats to the validity of this study.

\begin{itemize}
    
    \item \textbf{Selection Bias.} While platforms like LinkedIn, Reddit, and GitHub were options to recruit study respondents, we relied on the npm registry, as this ensured a more targeted approach to reach developers who are engaged in maintaining and developing packages for the npm ecosystem. Although this strategy improves relevance, it also limits the diversity of the participant pool, which could affect the generalizability of the results.
    
    \item \textbf{Generalizability Bias.} With only \SurveyParticipantCount completed responses, our sample may not represent the broader npm package developer population, limiting the generalizability of our findings.
    
    \item \textbf{Self-selection Bias.} While our study received responses from \SurveyParticipantCount npm package developers, there exists the possibility that these respondents have a strong interest in package security and may have been more likely to participate, which can skew the results.

    \item \textbf{Response \& Acquiescence Bias.}  Relying on self-reported data can lead to reporting biases, as respondents may exaggerate or underreport their experiences, and we cannot verify their claims. Additionally, there is a risk that differing interpretations of survey questions may lead to varied responses.
    
    \item \textbf{Subjectivity.} The classification of free-text responses inherently involves interpretation and potential subjectivity. To address this, two researchers independently analyzed the data and resolved differences through discussion. All team members brought extensive experience in qualitative methods and OSS, which further strengthened the reliability of the analysis.
    
    \item \textbf{Restrictive Question Formats.} The use of single-choice or multiple-choice questions can limit the depth of responses, as respondents may feel constrained by predefined options. To mitigate this, we included free-text fields or ``Other'' options.

    \item \textbf{Limited Scope of Questions.} The survey may overlook key areas of npm package security, leaving stakeholders without essential insights to improve security practices. 
        
    \item \textbf{Lack of Follow-Up.} Although the anonymous format of our survey promoted open and unbiased feedback from respondents, it hindered our ability to carry out follow-up interviews to clarify unclear responses and gather more detailed feedback.
    
    \item \textbf{Temporal Bias.} The findings reflect responses given at a specific time. Changes in technology, practices, or community concerns could make our results less relevant over time.

\end{itemize}

\section{Conclusion}
\label{sec:conclusion}

In this study, we explored how npm package developers approach security by collecting insights directly from practitioners. Through a survey of \SurveyParticipantCount developers, we uncovered important insights into how security is perceived, implemented, and prioritized within the npm ecosystem. We found that while security is widely valued, packages are often perceived only moderately secure, with supply chain attacks and dependency vulnerabilities seen as the most pressing threats. Practices are dominated by automated tools such as npm audit and Dependabot, whereas human-centered practices, such as security reviews and training, remain less common. Key barriers include time constraints, dependency complexity, and low trust in existing tools due to false positives. Furthermore, only about 40\% of respondents reported being satisfied with current npm security tooling. Looking ahead, developers call for improved detection tools, stronger protections for maintainers, better documentation, and cultural reinforcement through education and incentives. Our findings highlight actionable areas for improving tooling, ecosystem support, and secure software practices in OSS package management. Future work should explore longitudinal studies and cross-ecosystem comparisons (e.g., PyPI, Maven, RubyGems) to deepen understanding of security practices and identify ecosystem-specific versus generalizable solutions.

\bibliographystyle{ACM-Reference-Format}
\bibliography{references}

\end{document}